\documentclass[aps,pra,twocolumn,showpacs,superscriptaddress,nofootinbib,10pt]{revtex4-2}

\usepackage[utf8]{inputenc}
\usepackage[english]{babel}
\usepackage[T1]{fontenc}
\usepackage{csquotes}
\usepackage{physics}

\usepackage{amsmath,amsfonts}
\usepackage{bm}

\usepackage[dvipsnames]{xcolor}

\newcommand{\quantum}[0]{\textit{active}}
\newcommand{\classic}[0]{\textit{bath}}

\usepackage[colorlinks=true,linkcolor=black]{hyperref}
\usepackage[capitalise]{cleveref}

\usepackage{listings}

\usepackage{graphicx}
\graphicspath{{images/}}
\usepackage{qcircuit}

\newcommand{\angstrom}{\mbox{\normalfont\AA}}

\begin{document}
\title{Variational Embeddings for Many Body Quantum Systems}

\author{Stefano Barison}
\email{stefano.barison@epfl.ch}
\affiliation{Institute of Physics, \'{E}cole Polytechnique F\'{e}d\'{e}rale de Lausanne (EPFL), CH-1015 Lausanne, Switzerland}
\affiliation{Center for Quantum Science and Engineering, EPFL, Lausanne, Switzerland}
\affiliation{National Centre for Computational Design and Discovery of Novel Materials MARVEL, EPFL, Lausanne, Switzerland}

\author{Filippo Vicentini}
\affiliation{CPHT, CNRS, \'{E}cole polytechnique, Institut Polytechnique de Paris, 91120 Palaiseau, France}
\affiliation{Coll\`ege de France, Universit\'e PSL, 11 place Marcelin Berthelot, 75005 Paris, France}

\author{Giuseppe Carleo}
\affiliation{Institute of Physics, \'{E}cole Polytechnique F\'{e}d\'{e}rale de Lausanne (EPFL), CH-1015 Lausanne, Switzerland}
\affiliation{Center for Quantum Science and Engineering, EPFL, Lausanne, Switzerland}
\affiliation{National Centre for Computational Design and Discovery of Novel Materials MARVEL, EPFL, Lausanne, Switzerland}


\begin{abstract}

We propose a variational scheme to represent composite quantum systems using multiple parameterized functions of varying accuracies on both classical and quantum hardware.
The approach follows the variational principle over the entire system, and is naturally suited for scenarios where an accurate description is only needed in a smaller subspace.
We show how to include quantum devices as high-accuracy solvers on these correlated degrees of freedom, while handling the remaining contributions with a classical device.
We demonstrate the effectiveness of the protocol on spin chains and small molecules and provide insights into its accuracy and computational cost.

\end{abstract}

\maketitle

\begin{figure*}[ht]
    \includegraphics[width=1\textwidth]{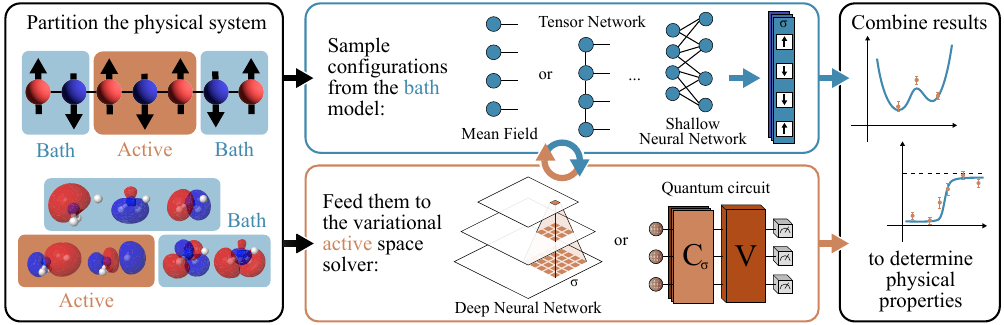} 
      \caption{Sketch of the hybrid algorithm. 
      We study a quantum system by locating a \classic \, and an \quantum \, partition.
      The model in the \classic \, partition is used to sample its physical configurations.
      Then, sampled configurations modify the output of the more accurate \quantum \, model, which can also be a quantum device.
      }
      \label{fig:emb_sketch}
\end{figure*}

\textit{Introduction} -- Predicting the properties of many-body quantum systems from accurate yet tractable simulations is one of the main goal of computational quantum physics.
While our computing capabilities have steadily grown over the years, we have to resort to approximated methods as the size of the systems to simulate increases \cite{szabo1996modern,KohnSham1965,Marzari2021}.
Several schemes replace interactions with an effective (mean) field generated by weakly-interacting components of the system, accurately describing only weakly interacting systems.  
However, many physical systems require treatment beyond mean-field to be described accurately \cite{Cohen2008_limitDFT}.

A strategy to overcome some of these limitations is to partition physical systems into multiple weakly interacting clusters.
Each cluster is then treated it with high-accuracy methods, while the interaction with the others -- the environment -- is introduced in a simplified way through a quantum \classic .
These approaches, called embedding schemes, circumvent the need to represent the entire system with uniform accuracy and allow us to use high-accuracy treatments on system sizes they cannot normally reach.
Embedding schemes have long been used in computational sciences, and have shown remarkable results in the study of physical and chemical quantum systems \cite{Georges96_DMFT,Knizia2012_dmet_hubbard,Knizia2013_dmet_mol}.
Their success stems from the natural partitioning a range of physical systems possess, such as molecular complexes ~\cite{Noodleman2002_nitro,Cady2008_emo_active,Kurashige2013}, or strongly interacting systems in a quantum bath \cite{Bockstedte2018,Gali2019_defects}.
However, the computational cost of the accurate method still limits the size of the relevant clusters, and how to systematically improve the results of those embedding schemes in size and accuracy remains unclear.

Variational approaches provide means to overcome these limitations.
They offer a way to systematically improve simulation results, achieving state-of-the-art accuracies while possessing a broad applicability ~\cite{wu2023variational}.
Employed since the early days of computational quantum mechanics \cite{McMillan_1965}, recently these class of methods have seen a surge in interest and application, mainly due to the introduction of machine learning models \cite{Carrasquilla2017,carleo_solving_2017,Wu2019,Vicentini_2019,Choo2020,Barrett2022,zhao2022scalable,Moreno2022_hidden,moreno2023enhancing,sinibaldi2023unbiasing,pescia2023messagepassing}.
Even so, a fully variational formulation of embedding schemes is still lacking in the literature.

The variational approach has also been introduced on quantum devices, and several successful experiments have been conducted ~\cite{Peruzzo_2014,Kandala_2017,Chiesa2019,Google2020_hf,arute2020observation,Neill_2021}.
However, due to limited resources in both the number of qubits and coherence times, treating on those devices entire physical systems of technological relevance seems unlikely in the near future ~\cite{childs2018_pnas,babbush2018_prx,yunseong2019_npjqi,motta2021_npjqi}.
Motivated by these limitations, several classical-quantum integrations have been proposed ~\cite{Bravyi2016_cutting,Tianyi2020_cutting,Mitarai_2021,piveteau2023circuit,Eddins_2022,Huembeli_2022, Shixin_2022,Shixin_2023}.
These quantum computing techniques are conceptually close to the embedding schemes presented above and show that, even at modest scales, quantum computers could provide insight into relevant physical problems \cite{Bauer2016,rubin2016hybrid,Rossmannek21_HFDFT, Rossmannek23_embedding}.
Therefore, considered their potential scalability, it is desirable for any new embedding scheme to feature quantum devices as high-accuracy solvers.

In this manuscript, we present a strategy to embed variational methods of different accuracies and computational costs together into the same scheme.
The proposed method is completely variational and global, meaning that the parameters of both models are optimized simultaneously.
We show how to extend the method to include quantum circuits as high-accuracy solvers, combining them with a range of classical ansatzes, including machine learning models \cite{carleo_solving_2017}.
We demonstrate the algorithm on spin systems with more than $10^{3}$ spins, proving that the method is able to obtain a good accuracy compared to standard variational ansatzes, while reducing the computational costs. 
Finally, we study small molecular systems, where the quantum device is used to model a selected active space and the mixed ansatz is modified to conserve the total number of fermions.

\vspace{3pt}
\textit{Variational mixed ansatz} -- Consider a physical system governed by a Hamiltonian $\hat{H}$ acting on the Hilbert space $\mathcal{H}$.
Without loss of generality, we can partition every Hilbert space $\mathcal{H}$ into two subspaces that we call \textit{active} $\mathcal{H}_A$ and \textit{bath}  $\mathcal{H}_B$, such that $\mathcal{H} = \mathcal{H}_A \otimes \mathcal{H}_B$.

This partitioning naturally induces the following rewriting of the Hamiltonian 
\begin{equation}
\label{eq:hami_mix}
    \hat{H} = \hat{H}_A \otimes \mathbb{I}_B + \mathbb{I}_A \otimes \hat{H}_B + \hat{H}_{\text{int}} = \sum_{j} \hat{H}^{A, j}\otimes \hat{H}^{B, j} \, ,
\end{equation}
where $\hat{H}_A$ ($\hat{H}_B$) contains all the terms of $\hat{H}$ acting non-trivially on the \textit{active} (\textit{bath}) sub-space, while $\hat{H}_{\text{int}} = \sum_{j} \hat{H}^{A, j}_{\text{int}} \otimes \hat{H}^{B, j}_{\text{int}}$ contains all the terms with nontrivial action on both sub-spaces.
We note that the physical system is separable if the interaction term $\hat{H}_{\text{int}}=0$.

To define the variational mixed model, we first consider a basis set for the bipartite Hilbert space $\{\ket{\bm{\sigma} , \bm{\eta}}\}$, where $\bm{\sigma} = (\sigma_1 , \dots , \sigma_{N_A})$ and $\bm{\eta} = (\eta_1 , \dots , \eta_{N_B})$ label the \textit{active} and \textit{bath} degrees of freedom, respectively.
Then, we write a quantum state in this basis as
\begin{equation}
\label{eq:mixed_model}
\ket{\Psi} =  \sum_{\bm{\sigma} , \bm{\eta} } \Psi(\bm{\sigma}, \bm{\eta}) \ket{\bm{\sigma}, \bm{\eta}}=\sum_{\bm{\sigma} , \bm{\eta} }  \alpha(\bm{\sigma}|\bm{\eta}) \, \beta(\bm{\eta}) \ket{\bm{\sigma}, \bm{\eta}} \, ,
\end{equation}
where we have decomposed the complete wave function of the quantum system into the product of two contributions.
The variational ansatz $\Psi_{\bm{\theta},\bm{\delta}}(\bm{\sigma}, \bm{\eta})$ is therefore obtained as the product of the two different ansatzes  $\alpha_{\bm{\theta}}(\bm{\sigma}|\bm{\eta}) \, \beta_{\delta}(\bm{\eta})$ depending on a set of parameters $\{ \bm{\theta},\bm{\delta} \}$.
One notable feature of the ansatz is that $\alpha$ has an explicit dependence on the \classic \, configurations $\bm{\sigma}$ in order to forge entanglement between this and the \quantum \, partition.
If $\alpha$ does not depend on $\bm{\sigma}$, the ansatz reduces to $\sum_{\bm{\sigma}} \alpha_{\theta}(\bm{\sigma}) \ket{\bm{\sigma}} \otimes \sum_{\bm{\eta} } \beta_{\delta}(\bm{\eta}) \ket{\bm{\eta}}$,  describing a factorized state.
The division allows to choose different variational wave functions with accuracies and computational costs tuned to the partition of the Hilbert space, \quantum \, or \classic, they model.

\begin{figure*}[ht]
    \includegraphics[width=1\textwidth]{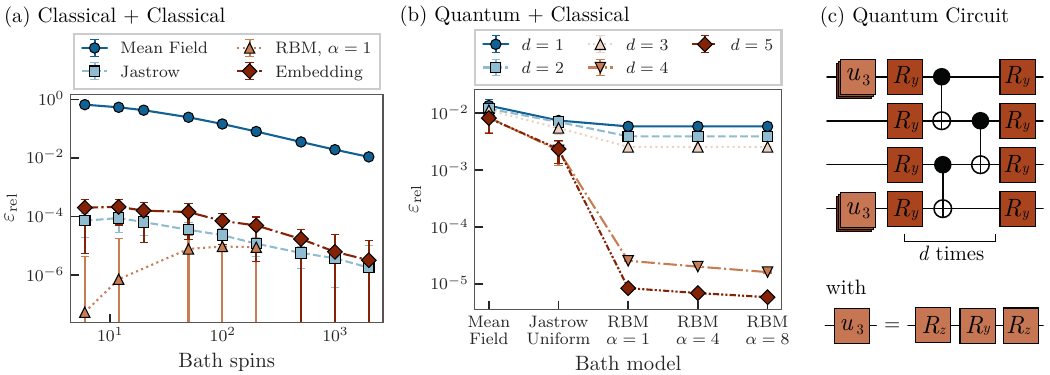} 
      \caption{
      	(a) Relative error $\varepsilon_{\text{rel}} = |E_{\text{var}}-E_{0}|/ E_{0}$ of the variational energy $E_{\text{var}}$ wrt to reference ground state energy $E_0$ as a function the number of spins in the \classic \, partition.
	    The number of spins in the \quantum \, partition is fixed at 3.
        The mixed model is obtained embedding a statevector for the \quantum \, subspace inside a uniform Jastrow ansatz for the \classic.
        Error bars indicate the statistical error of Monte Carlo sampling.
        (b) Relative error $\varepsilon_{\text{rel}}$ for different classical variational ansatzes and quantum circuit depths $d$. 
        The chain has 10 spins in the \classic \, partition and 3 in the \quantum \, one.
        We consider an ideal quantum simulator, therefore the error bars indicate the statistical error given from the classical sampling.
        (c) Sketch of the quantum circuit used in (b)}
      \label{fig:ising_composite}
\end{figure*}

The set of parameters can be optimized to represent the ground state of a quantum system, following a Variational Monte Carlo (VMC) approach \cite{McMillan_1965}.
The expectation value of the Hamiltonian is evaluated as 
\begin{multline}  
\label{eq:vmc_expect}
E = \frac{\bra{\Psi}\hat{H} \ket{\Psi}}{\braket{\Psi}} = 
\\ 
= \sum_{\bm{\sigma}, \bm{\eta} }\frac{|\Psi(\bm{\sigma},\bm{\eta})|^2}{\sum_{\bm{\sigma}'',\bm{\eta}''}|\Psi(\bm{\sigma}'',\bm{\eta}'' )|^2} \sum_{\bm{\sigma}' , \bm{\eta}'} \frac{\Psi(\bm{\sigma}',\bm{\eta}') }{\Psi(\bm{\sigma}, \bm{\eta})} \hat{H}^{\bm{\sigma}, \bm{\sigma}'}_{\bm{\eta}, \bm{\eta}'} 
\\
=  \sum_{\bm{\sigma}, \bm{\eta}} p_{\Psi}(\bm{\sigma},\bm{\eta}) E_{\text{loc}}(\bm{\sigma}, \bm{\eta}) \, ,
\end{multline}
where $\hat{H}^{\bm{\sigma}, \bm{\sigma}'}_{\bm{\eta}, \bm{\eta}'} = \bra{\bm{\sigma},\bm{\eta}} \hat{H} \ket{\bm{\sigma}',\bm{\eta}'}$ are the matrix elements of the operator $\hat{H}$ in this basis, $p_{\Psi}(\bm{\sigma},\bm{\eta}) = \frac{|\Psi(\bm{\sigma},\bm{\eta})|^2}{\sum_{\bm{\sigma}'',\bm{\eta}''}|\Psi(\bm{\sigma}'',\bm{\eta}'' )|^2}$ is the probability distribution associated to $\Psi$, and 

\begin{equation}
    \label{eq:Eloc}
    E_{\text{loc}}(\bm{\sigma}, \bm{\eta}) = \sum_{\bm{\sigma'} , \bm{\eta'}}  \frac{\Psi(\bm{\sigma'},\bm{\eta'}) }{\Psi(\bm{\sigma}, \bm{\eta})} \hat{H}^{\bm{\sigma}, \bm{\sigma'}}_{\bm{\eta}, \bm{\eta'}}
\end{equation}
is the local energy.
This suggests that the energy can be estimated by taking the sample mean of $E_{\text{loc}}(\bm{\sigma}, \bm{\eta})$ on a set of polynomially many samples $\bm{\sigma},\bm{\eta}$.
The samples can be generated from $p_{\Psi}$ using the Metropolis-Hastings algorithm \cite{Metropolis_1953}, or via direct exact sampling if the associated ansatz allows for it \cite{Sharir2020PRLAutoregressive}.

If we now make the extra assumption that 
\begin{equation}
\label{eq:active_norm}
\sum_{\bm{\sigma}} |\alpha_{\theta}(\bm{\sigma}|\bm{\eta})|^2 = 1 \, \, \, \,  \forall \bm{\eta}
\end{equation}
the energy estimation in \Cref{eq:vmc_expect} reduces to
\begin{equation}  
\label{eq:mixed_expect}
E = \sum_{\bm{\eta}} p_{\beta}(\bm{\eta}) \sum_{\bm{\sigma}} |\alpha(\bm{\sigma} | \bm{\eta} )|^2 E_{\text{loc}}(\bm{\sigma}, \bm{\eta}) \, ,
\end{equation}
indicating that the energy can be estimated by sampling $\bm{\eta} \sim p_\beta$ and evaluating the weighted local energy $\sum_{\bm{\sigma}} |\alpha_{\theta}(\bm{\sigma} | \bm{\eta} )|^2 E_{\text{loc}}(\bm{\sigma}, \bm{\eta})$.
We note that the sum over $\sigma$ scales exponentially in $N_A$, therefore can be performed only if the \quantum \, partition is small enough.
However, this strategy avoids to sample from the \quantum \, partition, and has to be preferred in those subsystem in which Markov Chain Monte Carlo can suffer from critical slowing down.

\vspace{3pt}
\textit{Embedding quantum devices} --
The assumption in \Cref{eq:active_norm} applies even if the \quantum \, partition is modelled using a variational quantum circuit.
Indeed, we can consider $\alpha_{\theta}(\bm{\sigma} | \bm{\eta} )$ as the complex amplitude associated to $\ket{\bm{\sigma}}$ of a state $U_{\theta}(\bm{\eta}) \ket{0}$ prepared using a quantum device, namely
\begin{equation}
    \label{eq:quantum_active_model}
    \alpha_{\theta}(\bm{\sigma}| \bm{\eta}) \ket{\bm{\sigma}} = \bra{\bm{\sigma}} U_{\theta}(\bm{\eta}) \ket{0} \ket{\bm{\sigma}} \, .
\end{equation}
Using the equality $\sum_{\sigma} \dyad{\bm{\sigma}} = \mathbb{I}$, when the \quantum \, subspace is modelled on a quantum device we can rewrite the state in \Cref{eq:mixed_model} as
\begin{equation}
    \label{eq:mixed_model_quantum}
    \ket{\Psi} =  \sum_{\bm{\eta} } \beta_{\delta}(\bm{\eta}) \ket{\bm{\eta}} \otimes U_{\theta}(\bm{\eta})\ket{0} \, .
\end{equation}

To measure the \quantum \, contribution on the quantum device, we use the decomposition of the Hamiltonian in \Cref{eq:hami_mix} and rewrite the energy evaluation as

\begin{equation}
    \label{eq:e_mix_quantum}
    E = \sum_{j ,\bm{\eta}} p_{\beta}(\bm{\eta}) \sum_{\bm{\eta}'} \frac{\beta_{\delta}(\bm{\eta}')}{\beta_{\delta}(\bm{\eta})} \bra{0}U_{\theta}^{\dagger}(\bm{\eta})  \hat{H}^{A,j} U_{\theta}(\bm{\eta}') \ket{0} H^{B,j}_{\bm{\eta},\bm{\eta}'} \, ,
\end{equation}

where $\bra{0}U_{\theta}^{\dagger}(\bm{\eta})  \hat{H}^{A,j} U_{\theta}(\bm{\eta}') \ket{0}$ is the overlap to evaluate on quantum hardware.
If $\bm{\eta}=\bm{\eta}'$, this reduces to the measurement of $\hat{H}^{A,j}$ on the circuit $U_{\theta}(\bm{\eta})$.
In the other case, the two unitaries $U$ are different and we have to resort to more general schemes, such as the Hadamard test \cite{Cleve_1998}.
More details on the evaluation of the quantum term are given in \cref{app:had_test}. 
From \cref{eq:e_mix_quantum} it is possible to show that the gradient w.r.t. the variational parameters of an observable $\hat{H}$ can be estimated as a classical expectation value over the same distribution $p_\beta$.
In \cref{app:hybrid_gradient} we provide a detailed discussion of quantum and classical gradient calculation.
We highlight that every step of this procedure, sketched in \Cref{fig:emb_sketch}, can be accomplished efficiently in polynomial time with respect to the system size.

\vspace{3pt}
\textit{Transverse Field Ising} -- As a first benchmark for our hybrid algorithm, we consider the problem of finding the ground state of a non-homogeneous Transverse Field Ising Model on a chain with open boundary conditions: 

\begin{equation}
\label{eq:tfim}
H_{\text{Ising}} =  \sum_{i=1}^{N-1} J_i \sigma^{z}_{i}\sigma^{z}_{i+1} + \sum_{i=1}^{N} \sigma^{x}_{i} \, .
\end{equation}

The first term accounts for interactions between spins, while the latter represents a local magnetic field along the transverse direction $x$.
We focus on bipartite systems where the \quantum \, partition has $J_i> 1$ and $J_i \ll 1$ in the \classic \, one.
Interaction between partitions is  $0<J_i<1$ and determines how entangled the two subsystems are.
In the \classic \, partition the spins interact only weakly between each others.
Therefore, a mean field ansatz or a uniform Jastrow \cite{Becca_Sorella_2017} provides a good approximation for this subspace.
However, these models are not as accurate in  the \quantum \, partition, due to the strong interactions.
For this reason, we create an hybrid model combining a statevector for the \quantum \, partition with a uniform Jastrow for the \classic .
The dependence of $\alpha$ upon $\bm{\eta}$  is introduced via a feed-forward neural network (FFNN). 
A detailed explanation of the hybrid models can be found in \cref{app:model}.

To test this model, we consider $N_A =3$ spins in the \quantum \, partition and vary the \classic \, spin number.
In this setting the total number of parameters of the hybrid model is 40, regardless of systems size.
In the panel (a) of \cref{fig:ising_composite} we compare the accuracies of different classical variational ansatzes with our hybrid approach as a function of the total number of spins.
We measure the relative error $\varepsilon_{\text{rel}} = |E_{\text{var}}-E_{0}|/ E_{0}$, where $E_{\text{var}}$ is the energy of the optimized model.
The ground state energy $E_0$ is obtained exactly for systems up to 30 spins, while we consider Matrix Product States optimized with the Density Matrix Renormalization Group algorithm (DMRG, \cite{White_1992}) for the bigger ones.
While more accurate models might be preferable at small system sizes, as the \classic \, subspace scales optimizing them becomes unfeasible, especially with higher-order optimizers \cite{Sorella_SR07}, as they have tens of thousands of parameters.
More details about the choice of the systems and model optimization can be found in \cref{app:ising_details} and \cref{app:opt_details}.

Then, we fix the \classic \, spin number $N_B=10$ and we explore the interplay between the quantum circuits and the classical models.
In the \quantum \, sub-space, we employ a hardware-efficient circuit with alternating layers of single-qubit $R_y$ rotations and two-qubit CNOT gates.
These circuits are combined with a mean field ansatz, a uniform Jastrow with nearest and next-to-nearest neighbors interaction, and  Restricted Boltzmann Machines (RBMs) Neural Quantum State \cite{carleo_solving_2017}.
Panels (b) and (c) of \cref{fig:ising_composite} show the results and the circuit used, respectively.

\begin{figure}
       \includegraphics[width=1\columnwidth]{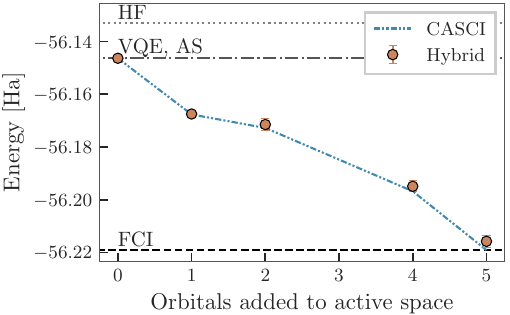} 
        \caption{
        Energy of the variational hybrid state approximating the ammonia molecule ground state as a function of the \classic \, orbitals considered.
        When an orbital is not in the variational subspace, its occupancy is frozen to be the one obtained with Hartree-Fock (HF).
        We start from VQE calculation in the active space (VQE, AS), then add the remaining orbitals using a Restricted Boltzmann Machine.
        The blue dash-dotted line represents the exact diagonalization results in the variational subspace (CASCI).}
        \label{fig:ammonia_ext}
\end{figure}

\vspace{3pt}
\textit{Molecular system} --  Next, we consider a molecular Hamiltonian in the Born-Oppenheimer approximation.
In the second quantization formalism, the Hamiltonian has the form
\begin{equation}
\label{eq:chem_hami}
H = \sum_{pq} h_{pq} a^{\dagger}_{p}a_{q} + \sum_{pqrs} V_{pqrs} a^{\dagger}_{p}a^{\dagger}_{q}a_{r}a_{s} \, ,
\end{equation}
where $a^{(\dagger)}$ are the fermionic annihilation (creation) operators, defined by the anticommutation relation $\{a^{\dagger}_{i},a_{j}\} = \delta_{ij}$, whereas $h_{pq}$ and $V_{pqrs}$ represent the one- and two-body integrals, respectively.
In this manuscript, we study the ammonia molecule (NH$_3$) using the Intrinsic Atomic Orbitals (IAO)\cite{Knizia2013iao} minimal basis set obtained from a mean-field calculation performed on the bigger aug-cc-pVQZ basis \cite{Kendall92_cc-orb}.
We consider a configuration in which one of the N$-$H bonds is stretched at a bond length of $1.5 ~\angstrom$. 
This stretching enhances the strong correlation in the electronic structure, as the atomic-like character of the constituent atoms is increased.
The \quantum \, space is represented by the Highest Occupied and the Lowest Unoccupied Natural Orbitals (HONO/LUNO) \cite{Barison2022iao}, for a total of four spin-orbitals.
We freeze the lowest energy orbital, corresponding to the $1s$ of the nitrogen atom.
Additional details regarding the ammonia simulation can be found in \cref{app:ammonia_details}.

We map the fermionic Hamiltonian of \cref{eq:chem_hami} onto a spin Hamiltonian using the Jordan-Wigner transformation ~\cite{Jordan93} (though other transformations could also be employed~\cite{Bravyi02_bk,Verstraete_2005,Whitfield_2016,Chen_2022}).
We constraint the ansatz explicitly in order to be particle-preserving.

We consider a variational quantum circuit in the \quantum \, subspace.
Starting from the result obtained using the Variational Quantum Eigensolver (VQE, \cite{Peruzzo_2014}) algorithm in this partition, we investigate the addition of the other orbitals as \classic .
The quantum hardware contribution is evaluated on a classical simulator.
As we report in \cref{fig:ammonia_ext}, when more orbitals are considered in the ansatz, the energy improves with respect to Hartree-Fock (HF \cite{szabo1996modern}) following the value obtained with exact diagonalization in that active variational space (Complete Active Space Configuration Interaction, CASCI \cite{szabo1996modern}).
As the size of the active space is increased, the CASCI value decreases towards the exact diagonalization result on the entire molecule (Full Configuration Interaction, FCI \cite{szabo1996modern}) with the frozen core approximation.
The quantum circuit includes one double excitation and two single excitation gates, resulting in 3 parameters to optimize.
Then, augmenting the ansatz with a RBM having real parameters, we are able to converge to FCI results.
More details about the hybrid model can be found in \cref{app:model}.


\vspace{3pt}
\textit{Discussion} -- In this manuscript, we have introduced a mixed approach that combines different variational embeddings to represent the ground state of interacting quantum systems from first principles.
The approach allows to exactly sample on the \quantum \, subspace if the corresponding ansatz is normalized, and allows to use quantum devices as \quantum \, solvers.
We have successfully tested our method on spin systems and molecular Hamiltonians, demonstrating its potential.

Many paths for followup research can be envisaged for the near future.
Alternative neural network architectures for representing quantum systems, beyond Restricted Boltzmann Machines \cite{Moreno2022_hidden,zhao2022scalable,Pfau2020_ferminet,pescia2023messagepassing}, can be explored.
Moreover, our scheme could also be used with minor modifications to study dynamical properties of quantum systems.

Under the quantum computation perspective, the effect of hardware noise on the results should be investigated.
Indeed, the addition of a classical model might provide some robustness during the optimization.
However, the importance of the method might extend even further in an age of fault-tolerant quantum computation, as offloading the treatment of weakly entangled partitions to a classical computer might still be advantageous.
Finally, as also pointed out in \cite{Mishmash2023}, strategies to optimally partition the physical system of interest into the \quantum \, and \classic \, subspaces are worth exploring.


\section*{Acknowledgments}

This research was supported by the NCCR MARVEL, a National Centre of Competence in Research, funded by the Swiss National Science Foundation (grant number 205602).
S.B. acknowledges  G. Gentinetta, C. Giuliani, F. Metz, A. Sinibaldi, S. Battaglia and J.R. Moreno for insightful discussions.

\section*{Data \& code availability}

Data and code are available on GitHub \cite{barison2023github}.
Simulations are performed using the open source libraries Netket \cite{Netket}, Pennylane \cite{pennylane}, and ITensor \cite{itensor}.
The molecular Hamiltonian is obtained with PySCF \cite{Sun2020recent} and Qiskit \cite{Qiskit}.
The colors used for the plots are part of the scientific colormap database \cite{Crameri2020}.


%



\appendix
\onecolumngrid


\section{The model}
\label{app:model}

In this section, we give a more detailed description of the mixed model.
As indicated in the main text, we consider a variational quantum state

\begin{equation}
	\label{eq:app_mixed_model}
\ket{\Psi} =  \sum_{\bm{\sigma} , \bm{\eta} } \Psi_{\theta,\delta}(\bm{\sigma}, \bm{\eta}) \ket{\bm{\sigma}, \bm{\eta}}=\sum_{\bm{\sigma} , \bm{\eta} }  \alpha_{\theta}(\bm{\sigma}|\bm{\eta}) \, \beta_{\delta}(\bm{\eta}) \ket{\bm{\sigma}, \bm{\eta}} \, ,
\end{equation}

where $\alpha_{\theta}$ and $\beta_{\delta}$ are two independent variational models describing the \quantum \, and \classic \, subspace, respectively.
We assume they are both computationally tractable \cite{Nest2009ProbabilisticMethods}, meaning that un-normalized complex amplitudes may be queried in polynomial time, and the Born probability amplitude may be sampled in polynomial time.
The product can be used as variational ansatz and we can write expectation values of physical observables $\hat{O}$ as
\begin{equation}
    \label{eq:app_vmc_expect}
    \frac{\bra{\Psi}\hat{O} \ket{\Psi}}{\braket{\Psi}} = \sum_{\bm{\sigma}, \bm{\eta} }\frac{|\Psi(\bm{\sigma},\bm{\eta})|^2}{\sum_{\bm{\sigma}'',\bm{\eta}''}|\Psi(\bm{\sigma}'',\bm{\eta}'' )|^2} \sum_{\bm{\sigma}' , \bm{\eta}'} \frac{\Psi(\bm{\sigma}',\bm{\eta}') }{\Psi(\bm{\sigma}, \bm{\eta})} \hat{O}^{\bm{\sigma}, \bm{\sigma}'}_{\bm{\eta}, \bm{\eta}'} =  \sum_{\bm{\sigma}, \bm{\eta}} p_{\Psi}(\bm{\sigma},\bm{\eta}) O_{\text{loc}}(\bm{\sigma}, \bm{\eta}) \, .
\end{equation}

Therefore, these expectation values can be estimated by sampling $\bm{\sigma},\bm{\eta}$ configurations from $p_{\Psi}(\bm{\sigma},\bm{\eta})= \frac{|\Psi(\bm{\sigma},\bm{\eta})|^2}{\sum_{\bm{\sigma}'',\bm{\eta}''}|\Psi(\bm{\sigma}'',\bm{\eta}'' )|^2}$ using the Metropolis-Hastings algorithm \cite{Metropolis_1953}, or other polynomially-efficient schemes, and evaluating the expectation value as 

\begin{equation}
    \label{eq:app_vmc_expect_mcmc}
    \frac{\bra{\Psi}\hat{O} \ket{\Psi}}{\braket{\Psi}} = \sum_{\bm{\sigma}, \bm{\eta}} p_{\Psi}(\bm{\sigma},\bm{\eta}) O_{\text{loc}}(\bm{\sigma}, \bm{\eta})  = \mathbb{E}_{\bm{\sigma}, \bm{\eta} \sim p_\Psi}\left[ O_{\text{loc}}(\bm{\sigma}, \bm{\eta}) \right]\, \sim \frac{1}{M} \sum_{\bm{\sigma}, \bm{\eta} \sim p_\Psi} O_{\text{loc}}(\bm{\sigma}, \bm{\eta}) \, ,
\end{equation}
where $M$ is the number of samples.
If we additionally assume that 
\begin{equation}
\label{eq:app_active_norm}
\sum_{\bm{\sigma}} |\alpha_{\theta}(\bm{\sigma}|\bm{\eta})|^2 = 1 \, \, \, \,  \forall \bm{\eta} \, ,
\end{equation}

the probability distribution $p_\Psi$ can be rewritten as

\begin{equation}
    \label{eq:app_beta_prob}
    p_{\Psi}(\bm{\sigma},\bm{\eta})= \frac{|\Psi(\bm{\sigma},\bm{\eta})|^2}{\sum_{\bm{\sigma}'',\bm{\eta}''}|\Psi(\bm{\sigma}'',\bm{\eta}'' )|^2} = \frac{|\beta(\bm{\eta})|^2}{\sum_{\bm{\eta}''}|\beta(\bm{\eta}'')|^2} |\alpha(\bm{\sigma}|\bm{\eta})|^2 = p_{\beta}(\bm{\eta}) |\alpha(\bm{\sigma}|\bm{\eta})|^2 \, ,
\end{equation}

indicating that expectation values can now be estimated as
\begin{equation}
    \label{eq:app_hybrid_expect_mcmc}
    \frac{\bra{\Psi}\hat{O} \ket{\Psi}}{\braket{\Psi}} = \sum_{\bm{\eta}} p_{\beta}(\bm{\eta}) \sum_{\bm{\sigma}} |\alpha(\bm{\sigma}|\bm{\eta})|^2 O_{\text{loc}}(\bm{\sigma}, \bm{\eta})  = \mathbb{E}_{\bm{\eta} \sim p_\beta} \left[ \tilde{O}_{\text{loc}}(\bm{\eta}) \right]\, \sim \frac{1}{M} \sum_{\bm{\eta} \sim p_\beta} \tilde{O}_{\text{loc}}(\bm{\eta}) \, ,
\end{equation}
where we introduced the weighted local observable $\tilde{O}_{\text{loc}}(\bm{\eta}) = \sum_{\bm{\sigma}} |\alpha(\bm{\sigma}|\bm{\eta})|^2 O_{\text{loc}}(\bm{\sigma}, \bm{\eta})$.
This implies that we can restrict our sampling to the \classic \, partition, and exactly compute the \quantum \, model contribution.
In the following, we will see that this is particularly useful when a quantum device is used to model the \quantum \, subspace.

\subsection{Classical models}
\label{subapp:classic_models}

In the study of spin and molecular systems presented in the main text, we considered many different classical models. 
Here we are going to list some of them

\begin{itemize}
    \item Mean Field ansatz: a product state ansatz of the form
    \begin{equation}
    \label{eq:app_mean_field}
        \psi(\bm{\sigma}) = \prod_{i=0}^{N} \frac{1}{\sqrt{1+ e^{-\lambda \sigma_i}}} \, ,
    \end{equation}
    where the single spin amplitude is modelled with a sigmoid and $\lambda$ is the only complex variational parameter.
    \item Uniform Jastrow ansatz: the wave function amplitude is 
    \begin{equation}
    \label{eq:app_jastrow_uniform}
        \psi(\bm{\sigma}) = \exp[ \sum_{i=0}^{N} J_1\sigma_{i}\sigma_{i+1} + J_2 \sigma_{i}\sigma_{i+2}] \, ,
    \end{equation}
    where $J_1$ and $J_2$ are the only two complex variational parameters, independent on the system size $N$, and periodic boundary conditions are applied, such that $\sigma_{N+1} \equiv \sigma_0$.
    \item Jastrow ansatz:
    \begin{equation}
    \label{eq:app_jastrow}
        \psi(\bm{\sigma}) = \exp[ \sum_{i=0}^{N} J^{1}_i\sigma_{i}\sigma_{i+1} + J^{2}_i \sigma_{i}\sigma_{i+2}] \, ,
    \end{equation}
    where now the ansatz contains two complex vectors of length $N$, for a total of $2N$ parameters.
    \item Restricted Boltzmann Machines (RBM): we used RBMs to model the wave function amplitudes as is \cite{carleo_solving_2017}.
    The wave function amplitude reads
    \begin{equation}
    \label{eq:app_rbm}
        \psi(\bm{\sigma}) = \exp[\sum_{j=0}^{N} a_j \sigma_j ] \times \prod_{i=0}^{\alpha N} \cosh{ \left[ b_i + \sum_{j=0}^{N} W_{ij}\sigma_j \right]}  \, ,
    \end{equation}
    where $\alpha N$ with $\alpha \in \mathbb{N}$ is the number of hidden degrees of freedom we trace out, and $\{ a, b, W\}$ are the complex variational parameters of the model.
    The total number of parameters is therefore $(\alpha +1) N + \alpha N^2$.
    \item State-vector: a vector of $2^N$ complex parameters, representing the wave function. 
    Can be used only for small system sizes.

\end{itemize}

In the following we illustrate how these different models are combined between them and with variational quantum circuits.

\subsection{Forging entanglement between models}
\label{subapp:backflow}

In the main text we highlight that the dependence of $\alpha(\bm{\sigma} | \bm{\eta})$ upon $\bm{\eta}$ is central to describe entanglement between the two subsystems.
In this section we focus on how we designed this dependence in the models we used on spin and molecular systems.

\begin{equation}
\label{eq:app_active_bf1}
\alpha_{\theta}(\bm{\sigma} | \bm{\eta}) = \alpha_{\theta}\left[\bm{\sigma}( \bm{\eta})\right] \, \, \text{where} \, \,  \bm{\sigma}( \bm{\eta}) = \bm{\sigma}  + f( \bm{\eta})
\end{equation}

This is very close to what is defined as \textit{backflow transformation} \cite{szabo1996modern,Becca_Sorella_2017} for the description of interacting fermionic systems.
In that case, correlation between particles is introduced in a single Slater determinant wavefunction making each of their coordinates a function of the other ones.
Instead, in this case we are making \quantum \, space configurations a function of the \classic \, space ones.

The backflow function $f$ is in general unknown, therefore it is a good practice to model it with a variational ansatz $f_{\theta '}$, where $\theta '  \in \theta$ is the subset of parameters of $\alpha$ parameterizing the backflow function.
Using the chain rule, the gradient with respect to $\theta '$ parameters reads

\begin{equation}
\label{eq:app_grad_active_bf1}
    \nabla_{\theta'} \alpha\left[ \bm{\sigma} + f(\bm{\eta})\right] = \sum_{i} \frac{\partial \alpha\left[ \bm{\sigma} + f(\bm{\eta})\right]}{\partial f_i} \nabla_{\theta'} f_i(\bm{\eta})
\end{equation}

where $f_i$ runs over the dimension of the output, which may vary.
Another possibility is to modify the output of the \quantum \, model itself, as an example

\begin{equation}
    \label{eq:app_active_bf2}
    \alpha_{\theta}(\bm{\sigma} | \bm{\eta}) = \tilde{\alpha}_{\theta''}(\bm{\sigma})  + f( \bm{\eta}) \, ,
\end{equation}
where $\theta ''$ indicates the set of $\theta \notin \theta'$, with suitable normalization in case we want to satisfy \cref{eq:app_active_norm}.
The chain rule applies also in this case and allows us to compute gradients of the backflow function with standard automatic differentiation algorithms ~\cite{Baydin18_autodiff}.

\begin{figure*}[ht]
    \includegraphics{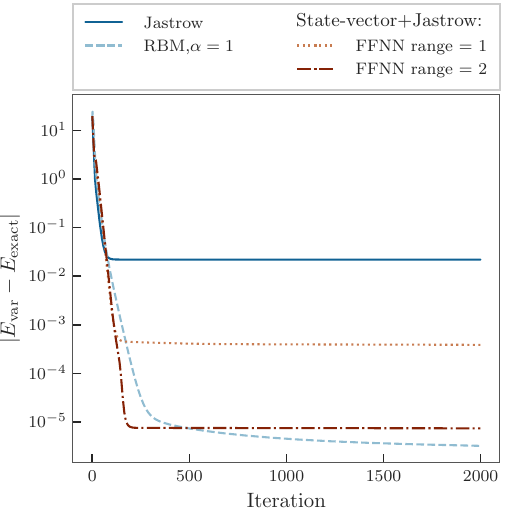} 
     \caption{Optimization of different classical variational ansatzes to represent the ground state of a Transverse Field Ising chain with 3 spins in the \quantum \, subspace and 8 in the \classic \, one. 
     The optimization is performed exactly without Monte Carlo sampling to avoid statistical noise. }
     \label{fig:ent_study}
\end{figure*}

In particular, in the study of the Transverse Field Ising Model we considered the combination of a uniform Jastrow ansatz modelling the \classic \, subspace with a State-vector for the \quantum \, one.
The backflow is realized with a Feed-Forward Neural Network (FFNN) that takes $\bm{\eta}$ as input ad outputs a complex vector in $\mathbb{C}^{2^{N_A}}$. 
Given that in the \classic \, subspace the spins are only weakly interacting between them, we restricted the input of the FFNN to a range of spins surrounding the active partition, which can be tuned in the model definition.
This allowed us to fix the embedded model total parameters independently to the system size, and to scale our simulations to $> 10^3$ spins.

In the following we will look at how the range of spins affects the final accuracy of the mixed model.
Referring to the notation presented in the main text, we consider a Ising chain with 3 spins and $J_{A}=10$ in the \quantum \, subspace and with 8 spins and $J_{B}=10^{-2}$ in the \classic \, one.
The interaction strength is $J_{\text{int}}=2$.
We compare the results of a Jastrow (22 parameters) and of an RBM (143 parameters) with an mixed ansatz composed by a Jastrow (16 parameters) and a State-vector (8 parameters).
The backflow function is a FFNN ad we fix its range to the nearest-neighbours spins (range $=1$, 38 parameters, 54 in total) and to the next nearest-neighbours (range $=2$, 68 parameters, 84 in total).
The results are reported in \cref{fig:ent_study}.

We see that the addition of the State-vector in the \quantum \, subspace improves the accuracy of two order of magnitudes at range $=1$ already.
When the range is increased, the final error is decreased by another 2 orders of magnitude, almost matching the accuracy of the RBM, which has almost the double of parameters.

\subsection{Quantum devices as high accuracy solvers}
\label{subapp:model_quantum}

In the main text we showed that the embedding scheme can be generalize to accommodate quantum devices as high accuracy solvers. In this case, we consider $\alpha_{\theta}(\bm{\sigma}| \bm{\eta}) \ket{\bm{\sigma}} = \bra{\bm{\sigma}} U_{\theta}(\bm{\eta}) \ket{0} \ket{\bm{\sigma}}$ and the ansatz reads

\begin{equation}
    \label{eq:app_mixed_model_quantum}
    \ket{\Psi} =  \sum_{\bm{\eta} } \beta_{\delta}(\bm{\eta}) \ket{\bm{\eta}} \otimes U_{\theta}(\bm{\eta})\ket{0} \, .
\end{equation}

We saw that expectation values of observables can be estimated using this ansatz, provided that is decomposed into a linear combination of separable terms, namely $\hat{O} = \sum_{k} \hat{O}_k=\sum_{k} c_k \hat{O}^{A}_{k}\otimes \hat{O}^{B}_{k}$, with $c_k \in \mathbb{C}$.
We remark that every operator admits such a decomposition.
For most physical observables (e.g. total magnetization, local occupation, dipole moments, \dots), the number of terms is polynomial in the system size.
The statistical error of the estimation of $\expval{\hat{O}}$ using $M_k$ samples is
\begin{equation}
\label{eq:app_var_mixed}
    \epsilon = \sqrt{\sum_k |c_k|^2 \text{Var}\left[\hat{O}_k\right] \big/ M_k}.
\end{equation}
Contrary to what happens in the context of standard Variational Monte Carlo, $\text{Var}\left[\hat{O}_k\right] \neq 0 $ even when the variational states get close to the eigenstates of $O$.

\begin{figure*}[t]
    \includegraphics[width=1\textwidth]{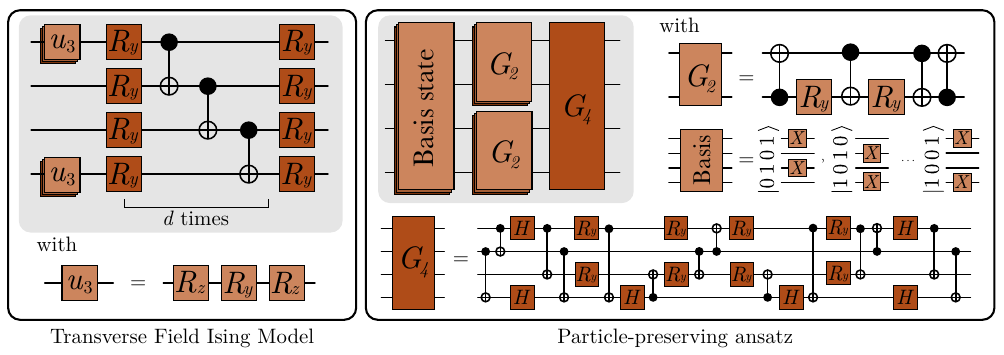} 
     \caption{Sketch of the variational circuits used as ansatzes for the \quantum \, subspace in the study of the Transverse Field Ising Model and the ammonia molecule. 
     The light orange gates that are repeated multiple times are controlled by the \classic \, configurations. }
     \label{fig:circ_sketch}
\end{figure*}

Finally, we discuss the structure of the parameterized quantum circuit.
This structure depends on the quantum system under study and the amount of quantum resources at disposal.
For the calculations on the Transverse field Ising Model we use a hardware-efficient ansatz consisting of a layer of $R_y$ rotations followed by a layer of CNOTs with linear connectivity, both repeated one or multiple times.
A scheme of the circuit can be found in \cref{fig:circ_sketch}.
The dependence of the circuit on the \classic \, subspace configurations is introduced via a Feed-Forward Neural Network (FFNN) that takes $\bm{\eta}$ as a input and outputs a set of six angles $\theta' \in \big[ 0, 2\pi )^{\times 6}$.
We will refer to this FFNN as a sample-to-angle function.
This set of angle determines the subset of classically-controlled rotations, which is composed of a rotation $u_{3}(\theta, \phi, \lambda) = R_z(\phi)R_y(\theta)R_z(\lambda)$ acting on the qubits encoding spins that have nearest neighbor interactions with spins that are in the \classic \, subspace.

To parameterize the ground state of the molecular system we implement a different, particle-preserving ansatz.
When no artificial constraint is imposed, only the total number of particles is conserved, but there is no such guarantee in each individual subspace.
In order to conserve the total number of particles in this mixed setting, we first restrict the RBM to sample only physical configurations by fixing a maximum and a minimum amount of electrons that may be present in the partition.
Then, the sample-to-angle function is extended to output the number of missing electrons in order to correctly initialize the quantum circuit.
Finally, we build the variational quantum circuit using only particle-preserving gates, in particular single and double excitation gates \cite{Anselmetti_2021,Arrazola2022universalquantum}.
If we now want to fix the number of particles in each subspace, we constrain the RBM to output only physical configurations with a precise number of electrons.
These modifications reduce the complexity of the problem and are readily extendable to bigger molecular systems.
A scheme of this circuit can be found in \cref{fig:circ_sketch}.
The actual implementation of the particle preserving gates will depend on the quantum hardware used for the experiment \cite{arute2020observation,Anselmetti_2021}.

\section{Calculating gradients of the hybrid ansatz}
\label{app:hybrid_gradient}

In this Section we show how to compute the gradient of expectation values with respect to the parameters of the mixed model.
Following the main text, given the mixed state $\ket{\Psi} =  \sum_{\bm{\sigma} , \bm{\eta} } \Psi_{\theta,\delta}(\bm{\sigma}, \bm{\eta}) \ket{\bm{\sigma}, \bm{\eta}}=\sum_{\bm{\sigma} , \bm{\eta} }  \alpha_{\theta}(\bm{\sigma}|\bm{\eta}) \, \beta_{\delta}(\bm{\eta}) \ket{\bm{\sigma}, \bm{\eta}}$ we define again the expectation value of a general operator $\hat{O}$ (of which the Hamiltonian $\hat{H}$ is a specific case) as

\begin{equation}
    \label{eq:app_vmc_expect2}
    \frac{\bra{\Psi}\hat{O} \ket{\Psi}}{\braket{\Psi}} = \sum_{\bm{\sigma}, \bm{\eta} }\frac{|\Psi(\bm{\sigma},\bm{\eta})|^2}{\sum_{\bm{\sigma}'',\bm{\eta}''}|\Psi(\bm{\sigma}'',\bm{\eta}'' )|^2} \sum_{\bm{\sigma}' , \bm{\eta}'} \frac{\Psi(\bm{\sigma}',\bm{\eta}') }{\Psi(\bm{\sigma}, \bm{\eta})} \hat{O}^{\bm{\sigma}, \bm{\sigma}'}_{\bm{\eta}, \bm{\eta}'} =  \sum_{\bm{\sigma}, \bm{\eta}} p_{\Psi}(\bm{\sigma},\bm{\eta}) O_{\text{loc}}(\bm{\sigma}, \bm{\eta}) \, .
\end{equation}

From the expression above, we derive the gradient with respect to $\theta$ and $\delta$ \cite{carleo_solving_2017}:

\begin{equation}
    \label{eq:app_gradient_classic}
    \nabla_{\theta,\delta} \langle \hat{O} \rangle = \sum_{\bm{\sigma}, \bm{\eta}}  p_{\Psi}(\bm{\sigma},\bm{\eta}) \left[ \nabla_{\theta}\log{\alpha_{\theta}(\bm{\sigma})} + \nabla_{\delta}\log{\beta_{\delta}(\bm{\eta})}\right] \left[ O_{\text{loc}}(\bm{\sigma}, \bm{\eta}) -  \langle \hat{O} \rangle \right] \, .
\end{equation}

As \cref{eq:app_vmc_expect2}, the gradient can be estimated by taking the sample mean of the expression in square brakets with samples generated from $p_{\Psi}$ using the Metropolis-Hastings algorithm \cite{Metropolis_1953}, or others sampling schemes.
If the \quantum \, subspace model is normalized as in \cref{eq:app_active_norm}, we can again reduce ourselves to sample only in the \classic \, partition, while exactly evaluating the \quantum \, contribution.
We highlight that all the machinery developed to optimize standard Variational Monte Carlo models, such as the stochastic reconfiguration protocol \cite{carleo_solving_2017}, can be used with the mixed models as well.
In the following, we will focus on the case in which the \quantum \, space is modelled on a quantum device, where extra caution has to be put in gradient evaluation.

\subsection{Gradient of the quantum-classical model}
\label{subapp:gradient_quantum}
For simplicity, we consider a classical ansatz $\beta$ which is holomorphic with respect to its  $n_c$ complex parameters $\delta \in \mathbb{C}^{n_c}$.
The unitaries $U_{\theta}$ defining the circuit, instead, have a set of $n_q$ real parameters $\theta \in \mathbb{R}^{n_q}$. 
We will now give the expression for the gradient of an expectation value with respect to the two different sets of parameters $\{\delta,\theta\}$.
We start again from the expression of the expectation value of $\hat{O}$ in \cref{eq:app_vmc_expect2}. For clarity, we consider the case of a separable observable the case $\hat{O} = \hat{O}^{A} \otimes \hat{O}^{B}$.
A non-separable observable will be then measured as a linear combination of separable terms.
With $\ket{\Psi} =  \sum_{\bm{\eta} } \beta_{\delta}(\bm{\eta}) \ket{\bm{\eta}} \otimes U_{\theta}(\bm{\eta})\ket{0} $ we obtain

\begin{equation}
    \label{eq:app_grad_explicit}
    \begin{cases}
    \begin{aligned}[b]
    \nabla_\delta \langle \hat{O} \rangle = \sum_{\bm{\eta}} p_{\beta}(\bm{\eta}) \bigg[ \sum_{\bm{\eta}'}  \hat{O}^{B}_{\bm{\eta} \bm{\eta}'}\frac{\beta_{\delta}(\bm{\eta}')}{\beta_{\delta}(\bm{\eta})}\hat{O}^{A}_{\bm{\eta} \bm{\eta}'} \nabla_{\delta} \log{\beta_{\delta}(\bm{\eta}')} - \langle \hat{O} \rangle \nabla_{\delta} \log{\beta_{\delta}(\bm{\eta})}  \bigg]
  \end{aligned} \\
    \nabla_\theta \langle \hat{O} \rangle = \sum_{\bm{\eta}} p_{\beta}(\bm{\eta}) \left[ \sum_{\bm{\eta}'} \hat{O}^{B}_{\bm{\eta} \bm{\eta}'} \frac{\beta_{\delta}(\bm{\eta}')}{\beta_{\delta}(\bm{\eta})} \nabla_{\theta} \hat{O}^{A}_{\bm{\eta} \bm{\eta}'} \right] \\
    \end{cases}
\end{equation}

where we indicated, for compactness, the matrix element on the classical partition $\langle \bm{\eta} | \hat{O}^B |\bm{\eta}' \rangle = \hat{O}^{B}_{\bm{\eta} \bm{\eta}'}$, the quantum expectation values $\bra{0} U^{\dagger}_{\theta}(\bm{\eta}) \hat{O}^A U_{\theta}(\bm{\eta}')\ket{0} = \hat{O}^{A}_{\bm{\eta} \bm{\eta}'}$.

Differently from the purely classical case, here we used also \classic \, models with real parameters $\delta\in\mathbb{R}^{n_c}$.
When this occurs, the first case of \cref{eq:app_grad_explicit} becomes

\begin{equation}
\label{eq:app_grad_local}
    \nabla_\delta \langle \hat{O} \rangle = \sum_{\bm{\eta}} p_{\bm{\eta}} \Bigg\{ 2\text{Re} \bigg[ \sum_{\bm{\eta}'}  \hat{O}^{B}_{\bm{\eta} \bm{\eta}'}\frac{\beta_{\delta}(\bm{\eta}')}{\beta_{\delta}(\bm{\eta})}\hat{O}^{A}_{\bm{\eta} \bm{\eta}'} \nabla_{\delta} \log{\beta_{\delta}(\bm{\eta}')}  - \langle \hat{O} \rangle \nabla_{\delta} \log{\beta_{\delta}(\bm{\eta})} \bigg]\Bigg\} \, .
\end{equation}

The second term of \cref{eq:app_grad_local}, on the other hand, represents the gradient evaluated on the quantum computer.
We evaluate this quantum term using an extension of the parameter shift rule \cite{schuld2018der,Mitarai_2018,parrish2019hybrid,mari2020estimating,banchi2020measuring} to the Hadamard test.
More explicitly, we can evaluate the derivative with respect to the $i$-th component of the parameter vector $\theta$ of the real and imaginary part of the overlap separately.
For the real part, we have

\begin{equation}
\label{eq:app_grad_quantum}
    \frac{\partial }{\partial \theta_i} \text{Re} \left[ \hat{O}^{A}_{\bm{\eta},\bm{\eta}'}(\theta) \right] = \frac{1}{2}\Bigg\{ \text{Re} \left[ \hat{O}^{A}_{\bm{\eta},\bm{\eta}'}(\theta+\frac{\pi}{2}e_i) \right] - \text{Re} \left[ \hat{O}^{A}_{\bm{\eta},\bm{\eta}'}(\theta-\frac{\pi}{2}e_i)\right] \Bigg\}
\end{equation}

where we indicated  $\hat{O}^{A}_{\bm{\eta},\bm{\eta}'}(\theta) = \bra{0}U_{\theta}^{\dagger}(\bm{\eta}) \hat{O}_A U_{\theta}(\bm{\eta}')\ket{0}$ to emphasize the $\theta$ dependence, and the real (or imaginary) part is evaluated using the Hadamard test procedure presented in \cref{app:had_test}.
Repeating the same procedure for every component of the parameter vector we obtain an estimation of the quantum term of the gradient.

\subsection{Unbiased gradient estimator}
\label{subapp:gradient_unbiased}

Recently it has been shown that the estimator of the gradient can affected by a systematic statistical bias or exponential sample complexity when the wave function contains some (possibly approximate) zeros \cite{sinibaldi2023unbiasing}.
This scenario is likely to occur in ground-state calculations of fermionic systems, such as the molecular system we reported in the main text.
For this reason, we also implemented an unbiased estimator for the gradient of the classical models.
As an example, for the molecular system we rewrote the gradient in \cref{eq:app_grad_local} as

\begin{equation}
    \label{eq:app_grad_unbiased}
    \nabla_{\delta} \langle \hat{O} \rangle_{\text{unbiased}} = \sum_{\bm{\eta}} p_{\bm{\eta}} \, \Bigg\{ 2 \text{Re} \bigg[ \sum_{\bm{\eta}'}  \hat{O}^{B}_{\bm{\eta} \bm{\eta}'}\frac{\nabla_{\delta}\beta_{\delta}(\bm{\eta}')}{\beta_{\delta}(\bm{\eta})}\hat{O}^{A}_{\bm{\eta} \bm{\eta}'} - \langle \hat{O} \rangle \nabla_{\delta} \log{\beta_{\delta}(\bm{\eta})} \bigg] \Bigg\} \, .
\end{equation}

\section{Optimization details}
\label{app:opt_details}

After the gradient is estimated as showed in \cref{app:hybrid_gradient}, we choose a suitable classical optimizer to tune the variational parameters of the hybrid ansatz.

For the classical simulations showed in the panel (a) of Fig. 2 in the main text, we used the stochastic reconfiguration protocol detailed in ~\cite{carleo_solving_2017}.
This can be done both when a standard or a bath-restricted sampling is performed.
In the former case we used the standard implementation provided in NetKet \cite{Netket}, while in the latter we created a custom Quantum Geometric Tensor class that can be found in \cite{barison2023github}.
The learning rate has been set as $\xi = 0.001$ and the diagonal shift for the S matrix is $0.001$.
The optimization of every model has been performed for 5000 iterations for systems up to 500 spins, while for the bigger ones the optimization steps performed are 2500.
Expectation values are evaluated using $10^3$ Monte Carlo samples.

For the quantum-classical simulations showed in Figs. 2b and 3 we employed two different optimizers.
The classical models are again optimized using the stochastic reconfiguration protocol ~\cite{carleo_solving_2017}.
We use a learning rate $\xi \in \left[0.005,0.01\right]$ and a regularization factor for the S matrix of $0.001$.
On the other side, for the sample-to-angle neural network and the quantum circuit, we used the first-order optimizer ADAM \cite{kingma2014adam}, with default values for the hyperparameters and a starting learning rate $\xi = 0.01$.
The quantum overlaps are evaluated exactly on a quantum simulator, while every expectation value is estimated using $10^4$ Monte Carlo samples for the \classic \, model.
The RBMs for both the Ising model and the ammonia molecule are initialized from a random normal distribution with zero mean and a standard deviation of $0.01$.


\section{Evaluating overlaps on quantum hardware}
\label{app:had_test}

Here we show how to evaluate the quantum terms $\bra{0}U_{\theta}^{\dagger}(\bm{\eta}) \hat{O}_A U_{\theta}(\bm{\eta}')\ket{0}$.
As indicated in the main text, when $\bm{\eta}=\bm{\eta}'$, the evaluation of the overlap reduces to the estimation of the expectation value of $O_A$ on the circuit $U_{\theta}(\bm{\eta}) \ket{0}$. This is a routine operation on quantum computer and a central part of most variational algorithms.

On the other hand, when $\bm{\eta} \neq \bm{\eta}'$, the evaluation of the overlap needs extra care.
We assume that $\bm{\eta} \neq \bm{\eta}'$ implies $U_{\theta}(\bm{\eta}) \neq U_{\theta}(\bm{\eta}')$, even if this is not true in general.
Those overlaps are complex quantities, thus we have to perform an Hadamard test to evaluate its real and imaginary part \cite{Cleve_1998}.
In the following, we describe how to measure the real part; the same procedure applies for the imaginary part, with a slight modification. 

The Hadamard test requires the addition of an auxiliary qubit, which has to be prepared in the $\ket{+} = H\ket{0}$ state, where $H$ here is the Hadamard gate.
First, we use the auxiliary qubit as control to apply an anti-controlled $U_{\theta}(\bm{\eta})$ gate to our circuit, obtaining the state $\frac{1}{\sqrt{2}} \left(U_{\theta}(\bm{\eta}) \ket{0} \ket{0} + \ket{0}\ket{1}\right)$.
Then, we perform a controlled $U_{\theta}(\bm{\eta}')$ gate, using again the auxiliary qubit as the control.
Finally, we measure $X_{\text{aux}} \otimes O_A$ on the circuit, that gives the real part of the overlap.
A schematic representation of the procedure can be found below, for a system of 2 qubits.

\begin{figure}[ht]
    \centerline{
\Qcircuit @C=1em @R=1.5em {
    \lstick{\ket{0}}  & \gate{H}  & \ctrlo{1}                  & \ctrl{1}                      & \gate{H} & \meter  \\
    \lstick{\ket{0}}  & \qw       & \multigate{1}{U_{\theta}(\bm{\eta})}  & \multigate{1}{U_{\theta}(\bm{\eta}')}    & \qw      & \meter  \\
    \lstick{\ket{0}}  & \qw       & \ghost{U_{\theta}(\bm{\eta})}         & \ghost{U_{\theta}(\bm{\eta}')}           & \qw      & \meter 
        }
    }   
\label{fig:losch_circ}
\end{figure}

To measure the imaginary part of the overlap, the same procedure applies, with the auxiliary qubit initialised in $ \frac{1}{\sqrt{2}}\left( |0\rangle -i |1\rangle \right) = HS^{\dagger}|0\rangle$.

We highlight that the described procedure represent the Hadamard test in the most general form.
However, a great simplification occurs when using variational ansatzes where only a set of gates depend on the \classic \, space configuration \cite{Ying2017rte}. 

\section{Details of the Ising simulations}
\label{app:ising_details}

In this section we will present the details regarding the simulations on the Transverse Field Ising model presented in Fig.2 of the main text.

Starting from computations in (a), here we considered a chain with $N_A=3$ spins in the \quantum \, partition and various $N_B$ in the \classic \, one.
In the \quantum \, partition we set $J_A = 10$ and in the \classic \, $J_B=0.1$, while the coupling between the two partitions is $J_{\text{int}} = 0.1$.
The parameters are chosen so that the Mean Field and the uniform Jastrow ansatzes (see \cref{subapp:classic_models}) provide a good approximation of \classic \, partition, while the interaction between the two partition is small but not negligible.
The hybrid ansatz is obtained embedding a Statevector in a uniform Jastrow, with the backflow contribution parameterized using a FFNN, for a total of 40 parameters, not depending on the \classic \, space size.
The parameters of the other methods scale as indicated in \cref{subapp:classic_models}, therefore $2N$ for the Jastrow, $2N+N^2$ for the RBM with $\alpha=1$, and 1 for the mean field, where $N= N_A + N_B$ is the total number of spins.

We changed the setting for the computations in panel (b). 
In this case we set $J_A = 1, J_B=0.25$ and $J_{\text{int}} = 0.5$.
This choice of $J_A$ and $J_{\text{int}}$ proved to be the hardest for the quantum device.
The interaction term $J_B$ has been increased to highlight the difference between the choice of a Mean Field ansatz and a much more expressive one, such as a Restricted Boltzmann Machine.

\section{Details of the ammonia simulation}
\label{app:ammonia_details}

Here we will discuss the details of the ammonia simulation we have presented in the main text.
In the minimal basis set, each hydrogen atom contributes with a single atomic orbital, while we have $5$ orbitals for the nitrogen atom, for a total of $8$ orbitals ($16$ spin-orbitals) for the entire molecule NH$_3$.
In order to determine a local \quantum  \, space of orbitals,  we started considering the usage of the Intrinsic Atomic Orbitals (IAO) \cite{Knizia2013iao,Senjean2021iao}.
This construction yields a minimal basis set of orbitals which can exactly represent self-consistent mean-field wave functions obtained with a bigger basis set.
It has been shown that this basis can be successfully employed in the context of classical embedding methods in order to systematically improve the results as the size of the bath is increased \cite{Nusspickel2022}.
This is possibile because  unlike minimal bases, IAO bases are naturally embedded into larger basis of one-electron orbitals. 
Moreover, IAO have already been shown to improve results of quantum simulations on quantum hardware \cite{Barison2022iao}.

\begin{figure}[ht]
       \includegraphics[width=0.5\columnwidth]{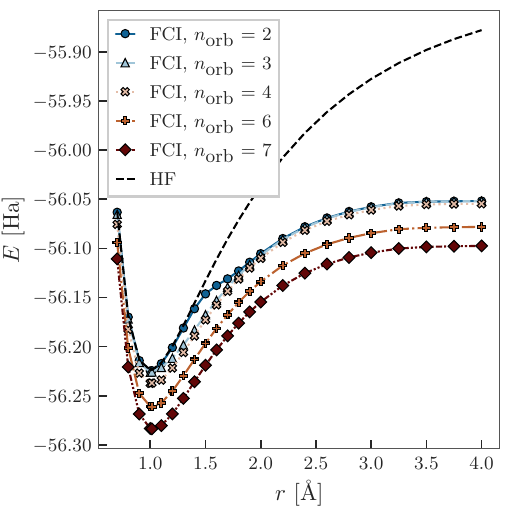} 
        \caption{Ground-state potential-energy curve of NH$_3$ along the NH$_3$ $\rightarrow$ NH$_2$ + H reaction path, using an IAO basis set obtained from a parent aug-cc-pVQZ mean field calculation. The plot reports the curves obtained with different active spaces, highlighting the role of each orbital in the qualitative description of the surface.}
        \label{fig:ammonia_pes}
\end{figure}

In order to construct the Hamiltonian operator we followed the procedure explained in \cite{Barison2022iao} using Qiskit \cite{Qiskit} and PySCF \cite{Sun2020recent}.
We considered a parent aug-cc-pVQZ basis ~\cite{Kendall92_cc-orb} on the NH$_3$ molecule (for a total of 218 orbitals) to perform the mean field calculation.
Given the Hartree-Fock molecular orbitals, we projected on the minimal basis set, obtaining a set of 8 orbitals polarized by the molecular environment.
Once we have the IAO set, we want to identify a subset of orbitals to be studied on the quantum device.
To this aim we performed a Moller-Plesset perturbation theory calculation (MP2 \cite{MP2_1934}) on the IAO minimal set, and computed the one-body reduced density matrix to obtain the natural orbitals.
Overall, the procedure prepares the set of orbitals $\{ \chi_k \}_{k=0}^{7}$ defined as

\begin{equation}
    \label{eq:app_IAO_transform}
    \chi_k = \sum_{j} U^{NO}_{kj}C^{IAO}_{ji} \phi_i \, ,
\end{equation}

where $C^{IAO}$ is the basis rotation defined by the IAO procedure, $U^{NO}$ is the unitary transformation to obtain the natural orbitals, and $\{ \phi_i \}_{i=0}^{7}$ is the original STO-6G minimal basis set.
Finally, we perform the frozen core approximation to reduce the number of orbitals to 7, and select as orbitals to be modeled on the quantum device the Highest Occupied Natural Orbital (HONO) and the Lowest Unoccupied Natural Orbital (LUNO). 

We consider the dissociation process of a single hydrogen atom from the ammonia molecule.
As can be see in \cref{fig:ammonia_pes}, the single determinant description of the wave function given by Hartree-Fock is not sufficient to give a potential energy surface in agreement with the exact diagonalization result.
Specifically, the quality of the Hartree-Fock approximation deteriorates when the N$-$H bond is greater than $1.5 ~ \angstrom$ long.
For this reason, we study the molecular configuration at $1.5 ~ \angstrom$ with the variational hybrid method,  choosing the two-orbitals (HONO/LUNO) active space to be studied on a quantum device and increasingly adding the other orbitals with the classical algorithm.

\subsection{Performance of VQE and VMC on the ammonia dissociation process}
\label{subapp:pure_computations}

\begin{figure}[ht]
       \includegraphics[width=0.5\columnwidth]{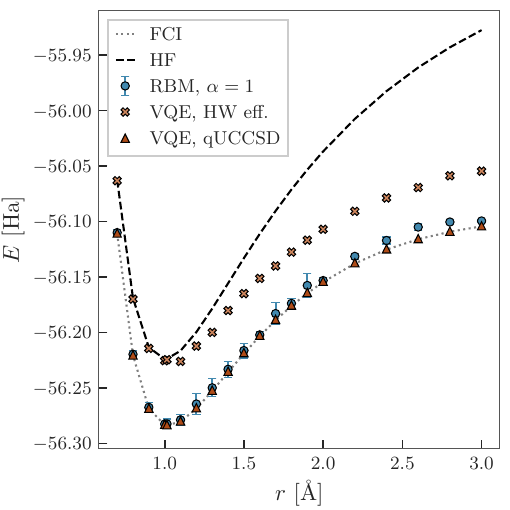} 
        \caption{Results of pure classical and quantum simulations of the ammonia dissociation process. The VQE calculations were performed on a classical simulator.}
        \label{fig:ammonia_pes_variational}
\end{figure}

In the main text we present a new method that combines quantum circuits and neural quantum states to prepare ground states of quantum systems.
We show how to use the method to study the ground state of the ammonia molecule (NH$_3$).
This system can be approached separately on a quantum computer using the Variational Quantum Eigensolver (VQE \cite{Peruzzo_2014}) and on a classical device using the Neural Network Quantum States.
Here we will show the performance of these two methods as a comparison.
We employ the RBM architecture with $\alpha=1$ as neural quantum state and use the exact sampler with $10^4$ samples per expectation value, similar to what we show in the main text.
With this choice of $\alpha$, the RBM has 224 complex parameters to optimize.
The parameters of the RBM are optimized using the Stochastic Reconfiguration protocol \cite{carleo_solving_2017} and 4000 optimization steps, apart from the points at $r > 2 ~ \angstrom$, which required 6000 steps.
The VQE calculations are performed using a classical simulator without shots and noise.
We consider the quantum Unitary Couple Cluster (qUCCSD) and a more near-term hardware friendly variational circuit as ansatz for the ground state wave function.
The results are presented in \cref{fig:ammonia_pes_variational}.

We see that the RBM architecture is able to represent accurately the ground state nearby the equilibrium, while the variational approximation  becomes less precise at large $r$.
We note that while we keep the architecture as close as possible to what we use in the main text, several different neural networks have been proposed to improve the simulation accuracy \cite{Barrett2022,Moreno2022_hidden,moreno2023enhancing}
On the other hand, the VQE with qUCCSD ansatz follows the FCI curve more closely, with optimizations converged in 100 steps starting from Hartree-Fock state.
However, the qUCCSD circuit is far from being suitable for near-term application.
Indeed, a fermionic system of 8 electrons in 14 spin-orbitals requires 24 singles and 180 double excitations to be included in a 14 qubit circuit.
For this reason we included the results of an hardware efficient ansatz, made of $4$ alternating layers of single qubit $R_y$ rotations and CNOT entangling gates (see main text), for a total of 70 single-qubit and 52 CNOT gates on nearest neighbour qubits.
With this hardware effcient ansatz the performance of the VQE deteriorates significantly, even if the potential-energy surface conserves a qualitative agreement with the FCI prediction.

The differences between the two ansatzes presented in \cref{fig:ammonia_pes_variational} highlight the role that the hybrid algorithm presented in the main text may have in extending the capabilities of current hardware.

\end{document}